# Automated detection and quantification of COVID-19 airspace disease on chest radiographs: A novel approach achieving radiologist-level performance using a CNN trained on digital reconstructed radiographs (DRRs) from CT-based ground-truth

Authors:

Eduardo Mortani Barbosa Jr.[1], Warren B. Gefter[1], Rochelle Yang[1], Florin C. Ghesu[2], Siqi Liu[2], Boris Mailhe[2], Awais Mansoor[2], Sasa Grbic[2], Sebastian Piat[2], Guillaume Chabin[3], Vishwanath R S.[4], Abishek Balachandran[4], Sebastian Vogt[5], Valentin Ziebandt[6], Steffen Kappler[7], Dorin Comaniciu[2]

[1] Perelman School of Medicine, University of Pennsylvania, Philadelphia, PA, USA
[2] Digital Technology and Innovation, Siemens Healthineers, Princeton, NJ, USA
[3] Digital Technology and Innovation, Siemens Healthineers, Paris, France
[4] Clinical Quality, Siemens Healthineers, Bangalore, Karnataka, India
[5] X-Ray Products, Siemens Healthineers, Malvern, PA, USA
[6] Imaging Decision Support, Siemens Healthineers, Erlangen, Germany
[7] XP Technology & Innovation, Siemens Healthineers, Forchheim, Germany

**Summary Statement:** A deep CNN trained using CT volumetric quantification of airspace disease projected onto digitally reconstructed radiographs can automatically quantify airspace disease on chest radiographs of patients with COVID-19 infection.

**Key Results:**

- Our novel approach leverages volumetric quantification of AD on CT in COVID-19 patients to better estimate the true error when performing AD quantification on CXR, via the intermediate step of constructing DRRs.

- Our DRR trained CNNs performed at least as well as expert human readers to quantify AD on CXR.

- This approach may increase efficiency and consistency in CXR interpretation of COVID-19 patients, while potentially providing prognostic disease biomarkers.

**Abbreviations:**

DRR = digitally reconstructed radiographs

CNN = convolutional neural network

CXR = chest radiograph

SARS-CoV-2 = severe acute respiratory syndrome coronavirus 2

COVID-19 = coronavirus disease 2019, caused by severe acute respiratory syndrome coronavirus 2

RT-PCR = reverse transcriptase – polymerase chain reaction

AI = artificial intelligence

AD = airspace disease, consisting of ground-glass opacities and consolidations

POa = Percentage of Opacity - Volume

POv = Percentage of Opacity - Area

MAE = mean absolute error

**Abstract**

Purpose: To leverage volumetric quantification of airspace disease (AD) derived from a superior modality (CT) serving as ground truth, projected onto digitally reconstructed radiographs (DRRs) to: 1) train a convolutional neural network to quantify airspace disease on paired CXRs; and 2) compare the DRR-trained CNN to expert human readers in the CXR evaluation of patients with confirmed COVID-19.

Materials and Methods: We retrospectively selected a cohort of 86 COVID-19 patients (with positive RT-PCR), from March-May 2020 at a tertiary hospital in the northeastern USA, who underwent chest CT and CXR within 48 hrs. The ground truth volumetric percentage of COVID-19 related AD (POv) was established by manual AD segmentation on CT. The resulting 3D masks were projected into 2D anterior-posterior digitally reconstructed radiographs (DRR) to compute area-based AD percentage (POa). A convolutional neural network (CNN) was trained with DRR images generated from a larger-scale CT dataset of COVID-19 and non-COVID-19 patients, automatically segmenting lungs, AD and quantifying POa on CXR. CNN POa results were compared to POa quantified on CXR by two expert readers and to the POv ground-truth, by computing correlations and mean absolute errors.

Results: Bootstrap mean absolute error (MAE) and correlations between POa and POv were 11.98% [11.05%-12.47%] and 0.77 [0.70-0.82] for average of expert readers, and 9.56%-9.78% [8.83%-10.22%] and 0.78-0.81 [0.73-0.85] for the CNN, respectively.
3
**Abstract**

Purpose: To leverage volumetric quantification of airspace disease (AD) derived from a superior modality (CT) serving as ground truth, projected onto digitally reconstructed radiographs (DRRs) to: 1) train a convolutional neural network to quantify airspace disease on paired CXRs; and 2) compare the DRR-trained CNN to expert human readers in the CXR evaluation of patients with confirmed COVID-19.

Materials and Methods: We retrospectively selected a cohort of 86 COVID-19 patients (with positive RT-PCR), from March-May 2020 at a tertiary hospital in the northeastern USA, who underwent chest CT and CXR within 48 hrs. The ground truth volumetric percentage of COVID-19 related AD (POv) was established by manual AD segmentation on CT. The resulting 3D masks were projected into 2D anterior-posterior digitally reconstructed radiographs (DRR) to compute area-based AD percentage (POa). A convolutional neural network (CNN) was trained with DRR images generated from a larger-scale CT dataset of COVID-19 and non-COVID-19 patients, automatically segmenting lungs, AD and quantifying POa on CXR. CNN POa results were compared to POa quantified on CXR by two expert readers and to the POv ground-truth, by computing correlations and mean absolute errors.

Results: Bootstrap mean absolute error (MAE) and correlations between POa and POv were 11.98% [11.05%-12.47%] and 0.77 [0.70-0.82] for average of expert readers, and 9.56%-9.78% [8.83%-10.22%] and 0.78-0.81 [0.73-0.85] for the CNN, respectively.

Conclusion: Our CNN trained with DRR using CT-derived airspace quantification achieved expert radiologist level of accuracy in the quantification of airspace disease on CXR, in patients with positive RT-PCR for COVID-19.

**MAIN MANUSCRIPT**

**Introduction**

Since its emergence in Wuhan, China in December 2019, the coronavirus SARS CoV-2 and its associated disease, COVID-19, has spread rapidly throughout the world. In March 2020 the World Health Organization (WHO) declared the COVID-19 outbreak a pandemic. COVID-19 has resulted in a global health and economic crisis, with exponential growth of cases despite substantial containment and mitigation measures (1,2).

As COVID-19 creates an unprecedented strain on health care systems, chest imaging (including chest radiographs (CXR) and chest CTs) has been increasingly utilized in suspected and confirmed cases (3,4). The clinical presentation of COVID-19 may vary from mild upper respiratory symptoms to severe dyspnea, fever and multifocal pneumonia, the latter the hallmark of severe disease. In the most severe cases, there is respiratory failure requiring admission to a critical care unit and mechanical ventilation (4,5).

The reference standard diagnosis of COVID-19 relies on the identification of the virus on nasopharyngeal swabs via reverse transcriptase-polymerase chain reaction testing (RT-PCR). However, a false negative rate of up to 30% has been reported, particularly in early disease (4-10). CXR and chest CT, therefore, may play an important role in the diagnosis of COVID-19, excluding other acute cardiopulmonary abnormalities and aiding in prognostication. Several studies have detailed the imaging features of COVID-19 pneumonia utilizing CXR and chest CT, with CT being commonly performed in China and South Korea for diagnosis and management (11-14).

The most typical chest CT feature of COVID-19 associated pneumonia is airspace disease (AD, consisting of ground glass opacities (GGO) with or without consolidations), often bilateral, multifocal, peripheral and lower lung zone predominant. This pattern resembles SARS, H1N1 influenza and cytomegalovirus infections more than other common lower respiratory viruses such as human metapneumovirus or respiratory syncytial virus (11-15). With disease progression, there may be increasing perilobular linear opacities and consolidation in a pattern characteristic of organizing pneumonia or, in the most severe cases, of diffuse alveolar damage (11, 13). CXR and chest CT are increasingly utilized to monitor disease progression in hospitalized patients with COVID-19 (16,17,18). Despite lower sensitivity and specificity than CT, CXR is often the first and only imaging study performed in COVID-19 patients, and likely will be increasingly utilized as the number of COVID-19 patients increases (18). This underscores the need to improve interpretation accuracy and quantification of airspace disease on CXR.

There have been studies using deep neural networks to make the CT reading efficient and reproducible by automatically segmenting COVID-19 related opacities and predicting quantitative biomarkers, such as the Percentage of Opacity (PO) (19, 20, 21). Although deep neural networks have also been used to aid the diagnosis of COVID-19 from CXR, most of the studies formulated the problem simply as an image-wise classification problem (22, 23, 24). Automatically quantifying the COVID-19 severity from CXR remains an open challenge, due to the difficulty of establishing high-confidence for training the neural networks.

Our goals were to assess: 1) the accuracy of COVID-19 AD quantification on CXR by an AI CNN against CT-derived AD volumetric quantification and AD area quantification projected on DRRs; and 2) to compare the performance of COVID-19 AD quantification by the AI CNN on CXR against expert human readers, using as ground truth AD volumetric quantification on CT.

**Materials and Methods**

<u>Patient and Image Selection</u>

This single institution, retrospective study obtained IRB approval with waiver of informed consent and was HIPAA compliant. We randomly selected 86 patients with the following inclusion criteria: positive RT-PCR for SARS-CoV-2 and a pair of CXR and chest CT performed within 48 hours of each other. All scans were obtained within the following date range: March – May 2020. Figure 1 details study design, inclusion and exclusion criteria. Table 1 details the demographics and clinical features of the cohort. CXR and CT images were de-identified using a standard anonymization profile in Sectra PACS and transferred through a secure file exchange to a computational cluster for imaging processing.

Expert Human Quantification of Airspace Disease (AD) on CXR

Two expert readers (subspecialty trained thoracic radiologists with 12 and over 30 years of experience) read each CXR independently, blinded to the results of the paired chest CT or any clinical information, except for being aware that each patient tested positive for SARS-CoV-2 by RT-PCR. Each reader segmented AD on a single frontal CXR using manual annotations on ITK-Snap [25].

CT 3D whole lung and 3D volumetric segmentation of AD

All CT datasets (86) were annotated manually by a trained annotator with two years of experience in annotating pulmonary abnormalities in chest CT. The annotations were supervised and revised by a radiologist with 6 years of experience in Chest CT. The CT readers were instructed to segment airspace disease (ground-glass opacities and consolidations). A subset of 13 outlier cases was further reviewed by 2 expert thoracic radiologists. ITK-Snap was utilized for manual AD segmentation [25].

Creation of DRRs and projection of CT-derived 3D volume quantification to 2D area quantification of airspace opacity

DRRs are generated as integrals over synthetic projection lines through the CT volume under a parallel projection geometry. The resolution of the resulting DRR is enhanced using a deep CNN, yielding isotropic resolution and reducing the average difference in resolution between DRR and typical chest radiographs (CXR). Finally, a frequency sub-band normalization is applied to reduce the noise and increase the contrast of the image. Figure 2 illustrates our method.

The whole lung segmentation and 3D AD segmentation in chest CTs are converted to volumetric masks, which can be projected in the 2D DRR in 2 different ways:

1. Anteroposterior Thickness Projection: For both whole lung and AD segmentation, the corresponding binary mask from CT can be projected using a line integral which measures the depth/thickness of the mask along the anterior-posterior axis under a parallel projection geometry.

2. Anteroposterior Intensity Projection: For both whole lung and AD segmentation, the corresponding binary mask from CT can be projected using a line integral over the intensity values of the CT image voxels within the mask along the anterior-posterior axis under a parallel projection geometry.

Segmentation of whole lungs on DRR and CXR

To train a lung segmentation neural network for the DRRs, we used the anteroposterior thickness projection derived from the 3D lung annotations in CT to establish the ground-truth annotations for DRR. A binarization of the projected mask is performed at a cut-off value of 38 mm. This value was established based on a qualitative assessment of optimal visibility of the lung boundary at different cut-off values. A deep convolutional segmentation network (26) is used to learn the mapping between input DRR and the established binary masks defining the area of the lungs. The trained network is also used for predicting the binary lung masks on the CXRs without further training.

Segmentation of COVID-19 airspace disease (AD) on DRR and CXR

We performed an intensity projection of the 3D ground truth mask describing the COVID-19 affected lung parenchyma. The resulting projection image threshold is set at 25,000 to obtain a binary mask for a given DRR. The value of the threshold is selected such that the average Mean Absolute Errors (MAE) between POv (ground-truth on CT) and POa on DRR is minimized. A deep convolutional neural network is used to learn the mapping between input DRR and the binary masks defining the area affected by airspace disease. The output of the system is represented as a pixel-wise probability map of airspace disease and constrained to the estimated area of the lung parenchyma. Given the selected threshold does not guarantee that every lesion quantified in the 3D-CT is visible in the DRR, there is a certain degree of label noise that can lead to a per-sample bias in the model estimation. We mitigated this limitation by training an ensemble of models and averaging the output of each to obtain an improved estimate (27). Similar to lung segmentation on CXR, the trained DRR lesion segmentation network is applied to the CXRs without further training. The processing time of the system to compute the POa, including segmenting both the lungs and the lesions, was 52 ms per CXR on an RTX 2080TI GPU. Figure 3 provides a visual illustration.

COVID-19 Airspaces Disease (AD) Severity Measures

The metrics used to assess the severity of airspace disease are detailed below.
Percentage of Opacity - Volume (POv): The POv is measured on CT scans and quantifies the percent volume of the lung parenchyma that is affected by airspace disease:

$$POv = 100 \times \frac{Volume\ of\ Airspace\ Disease}{Total\ Lung\ Parenchyma\ Volume}$$

Percentage of Opacity – Area (POa): The POa is measured on DRRs and CXRs and quantifies the percent area of the lung parenchyma that is affected by airspace disease:

$$POa = 100 \times \frac{Area\ of\ Airspace\ Disease}{Total\ of\ Lung\ Parenchyma\ Area}$$

The mean absolute error (MAE) conveys the difference between POa and POv for a pair of CXR and CT scan of the same patient acquired within 48 hours.

Statistical Analysis

We compared the POa values derived from the 2D annotations on CXR or DRR against the POv derived from the 3D CT annotations to calculate the MAE and Pearson correlation coefficients. All the MAE and the Pearson Coefficients were reported also with 95% bootstrapping (BS) with 1000 times resampling. The human expert POa were obtained using the average (Reader Avg.), the intersection (Reader Inter.) and the union (Reader Union) of the expert annotations. Both the single network CNN system (Sole System) and the ensemble CNN system (Ensemble CNN) were used for computing the POa automatically on the DRR and CXR images. The related sample t-test with one tailed p-values were computed to study the statistical significance ($p < 0.05$) between each pair of the expert POa results and the CNN system POa results. The statistical analysis was implemented in Python3.6.1 using the 'statsmodels' v0.12 package.

**Results**

Our cohort of 86 patients, who all had a positive confirmatory test (RT-PCR) for SARS-CoV-2, had a mean age of 59 years, with a wide range (25-93 yrs.) and balanced gender distribution. Of note, African Americans comprised 75% of the cohort. Almost half of the cohort were never-smokers (48%). We assessed outcomes to Jul 2020. The majority of the cohort recovered and has been discharged (72%) from the hospital, whereas 23% were deceased due to COVID-19 manifestations or related complications, and 5% remain in the hospital at the time of this writing. Table 1 details demographics and clinical features, Figure 1 details inclusion and exclusion criteria.

The training dataset used for generating the DRRs contains 1929 CT images obtained from multiple clinical sites. We also added 727 control CXR images to improve the system accuracy on control or mildly affected patients (Supplemental Table). A validation set consisting of 182 CT images was randomly chosen for model selection. The distribution of the POv (ground-truth POv) and POa (DRR POa) are summarized in Table 2. The MAE and the Pearson coefficient between the ground truth POa and POv are 5.27% and 0.97 respectively, which indicate intrinsic information loss from mapping 3D information into a parallel geometry planar coronal projection.

The two expert human readers demonstrated high inter-reader agreement (r = 0.817) for AD quantification on CXRs. Against the ground-truth of CT-derived POv, the MAE of the average human reader was 11.98% (13.15% and 12.14% for reader 1 and 2), with the correlation of 0.77 (0.70 and 0.76 for readers 1 and 2). The MAE of the reader intersection and union were 15.91% and 11.04%, with correlations 0.73 and 0.77. Figures 4, 5 and 6 demonstrate patient examples of AD quantification on CT, DRR and CXR, by expert readers and CNN.

As shown in Table 3, the single CNN quantified POa on CXR achieved MAE 9.78% and correlation 0.78 comparing to the ground truth POv derived from CT. With model ensemble, the MAE was further reduced to 9.56% while the correlation was increased to 0.81. Comparing to the CXR results, the CNN same systems achieved lower MAE (9.26% and 7.72%) and higher correlation (0.86 and 0.87) on DRR images acquired from the same patients within 48 hours. As shown in Table 4, statistical Student t test comparing the average, union and intersection of the CXR readers to the CNN demonstrated that the CNN overall performed slightly better than the expert human readers when compared to the POv ground-truth derived from CT, though the difference did not reach statistical significance.

**Discussion**

The rising number of COVID-19 cases will likely be paralleled by an increasing number of CXRs performed for diagnostic evaluation, underscoring the need to maximize CXR utility and value and reduce variability in interpretation in suspected and confirmed COVID-19 patients. While neither as sensitive nor as specific, CXRs will be utilized in much greater numbers than chest CTs, further supporting the need to augment CXR capability in the setting of the COVID-19 pandemic (28, 29). AD is the hallmark of pulmonary involvement in COVID-19, supporting the concept that AD quantification on CT and CXR carries diagnostic and potential prognostic implications (19, 30).

Quantitative assessment of extent of AD on CXR by deep CNN, as opposed to subjective or semi-quantitative evaluation by human readers, possesses three major strengths: 1) it provides disease quantification, not just binary output of disease present/absent, thereby carrying potential prognostic and management implications (18); 2) it increases consistency of AD evaluation over human readers, given high inter-reader and even intra-reader variability; 3) it can increase reading efficiency since most computational algorithms can generate results in a small fraction of the time a human reader would take to perform a similar task.

Most AI publications on COVID-19 imaging lack explicit quantification, instead focusing on classifying images based on the patient PCR status (positive vs negative) and utilizing primarily CT rather than much more widely available CXRs (31-35). Moreover, when CXRs have been used for quantification of airspace disease, the ground-truth has been human annotation, which we demonstrated is subjective, depends on considerable expertise and carries a non-negligible error, even when performed by highly experience expert thoracic radiologists.

Our work is innovative as it proposes leveraging a superior modality, i.e. CT, to provide a much more accurate ground-truth of AD quantification than can be obtained from CXR, to gauge the performance of human readers and deep CNN quantifying AD on less accurate CXRs. For that purpose, a crucial element is to project the 3D volume into a 2D coronal image that is as similar as possible to a CXR, which we labeled DRRs. This step introduces an intrinsic error due to the information loss resulting from the conversion of 3D volumetric AD (in CT) into 2D area AD (in DRR). Nonetheless, by knowing the ground-truth derived from CT, it is possible to obtain a binarization threshold on either the thickness or the intensity projection maps using the swipe search in the training dataset to minimize this intrinsic error. Without considering the image intensity information lost due to the projection, the intrinsic error was estimated to be at most 5.72% in our cohort. Though the CNNs were not explicitly optimized to output POv, the optimized training target helped lower the theoretical MAE bound between POa and POv for the CNN systems.

Our CNN, in particular, performed slightly better than the average of two highly trained expert human readers, with an MAE in the range of 9.56%, which is 3.84% higher than the estimated lower bound, although the difference with the average performance of human readers was not statistically significant. The Ensemble CNN was shown to outperform the single network with -0.2% MAE on the cohort, although the difference is not statistically significant. A clinically meaningful metric was the time required for the CNN to compute AD on each CXR, 52 ms per radiograph, versus several minutes for the human readers.

Our study has several limitations. Given that we selected a subset of patients who fulfilled multiple inclusion criteria (positive RT-PCR for SARS-CoV-2, and paired chest CT and CXR performed within 48 hours of each other) and due to the single center design, we have a relatively small sample size. The AD annotation on CT was performed by two human readers, without automated algorithms; however, outlier cases were secondarily reviewed by 2 additional expert radiologists and corrected. While every patient in the cohort had COVID-19 at the time the CXR and CT were obtained, it is possible that not all airspace disease detected was a manifestation of COVID-19. Our small sample lacked a control group who tested negative for SARS-CoV-2 by RT-PCR, precluding evaluation of diagnostic utility, which is a future research direction we intend to pursue.

In summary, we have devised a novel approach to improve the CXR quantification of AD in patients with COVID-19, leveraging quantification derived from a superior modality (CT) via the novel intermediate step of projecting the CT-derived AD mask into parallel projection coronal DRR. This approach provides a better ground-truth, with more accurate and quantitative understanding of the error accrued by both human readers and a CNN applied to CXR for AD 2D quantification. Furthermore, we showed that the CNN is at least as accurate as expert human readers for the task of CXR-based AD 2D quantification. Such a system, when deployed in a high-volume clinical setting, could substantially increase the consistency of interpretations and reduce reporting times, improving radiologist efficiency and throughput. Moreover, by providing quantitative measurements of AD that correlate with physiologic impairment, such a system could guide patient management and generate prognostic information. This is particularly true when applied longitudinally on serial CXRs, as it can quantify disease course over time. Future studies will evaluate the potential of this approach to automated quantification of AD in COVID-19 patients to serve as a prognostic imaging biomarker in predicting need for ICU admission and risk of development of ARDS.

AI algorithms that can accurately quantify COVID-19 associated airspace disease on CXR utilizing ground truth derived from a superior modality (CT) may improve diagnosis, management and prognostication in patients with COVID-19, thus augmenting the role of CXR in the COVID-19 pandemic.

**Acknowledgements:** We gratefully acknowledge the contributions of multiple frontline hospitals to this collaboration.

**Disclaimer:** The concepts and information presented in this paper are based on research results that are not commercially available

**References**

1. Novel Coronavirus (2019-nCov). World Health Organization. https://www.who.int/emergencies/diseases/novel-coronavirus-2019. Published March 29, 2020. Accessed July 24, 2020.

2. Situation Report - 69. World Health Organization. https://www.who.int/docs/default-source/coronaviruse/situation-reports/20200329-sitrep-69-covid-19.pdf?sfvrsn=8d6620fa_4. Published March 29, 2020. Accessed July 24, 2020.

3. Coronavirus Disease 2019 (COVID-19) - Evaluating and Testing PUI. Centers for Disease Control and Prevention. https://www.cdc.gov/coronavirus/2019-nCoV/hcp/clinical-criteria.html. Published March 29, 2020. Accessed July 24, 2020.

4. Wu Z, McGoogan JM. Characteristics of and Important Lessons From the Coronavirus Disease 2019 (COVID-19) Outbreak in China: Summary of a Report of 72 314 Cases From the Chinese Center for Disease Control and Prevention. *JAMA.* Published online February 24, 2020. doi:10.1001/jama.2020.2648.

5. Huang C, Wang Y, Li X, et al. Clinical features of patients infected with 2019 novel coronavirus in Wuhan, China. Lancet. 2020;395(10223):497-506. doi:10.1016/S0140-6736(20)30183-5

6. Xu XW, Wu XX, Jiang XG, et al. Clinical findings in a group of patients infected with the 2019 novel coronavirus (SARS-Cov-2) outside of Wuhan, China: Retrospective case series. BMJ. 2020;368(January):1-7. doi:10.1136/bmj.m606

7. Ai T, Yang Z, Hou H, et al. Correlation of Chest CT and RT-PCR Testing in Coronavirus Disease 2019 (COVID-19) in China: A Report of 1014 Cases. Radiology. 2020. Published Online February 26 2020. doi: 10.1148/radiol.2020200642

8. Tahamtan A. Real-time RT-PCR in COVID-19 detection: issues affecting the results. Expert Rev Mol Diagn. 2020; 20(5): 453-454. doi: 10.1080/14737159.2020.1757437

9. Long C, Xu H, Shen Q, et al. Diagnosis of the Coronavirus disease (COVID-19): rRT-PCR or CT? European Journal of Radiology. 2020; 126: 108961. doi: 10.1016/j.ejrad.2020.108961

10. He J, Luo L, Luo Z, et al. Diagnostic performance between CT and initial real-time RT-PCR for clinically suspected 2019 coronavirus disease (COVID-19) patients outside Wuhan, China. Respiratory Medicine. 2020; 168: 105980. doi: 10.1016/j.rmed.2020.105980

11. Rodriguez-Morales AJ, Cardona-Ospina JA, Gutiérrez-Ocampo E, et al. Clinical, laboratory and imaging features of COVID-19: A systematic review and meta-analysis. Travel Med Infect Dis. 2020;(February):101623. doi:10.1016/j.tmaid.2020.101623

12. Yoon SH, Lee KH, Kim JY, et al. Chest Radiographic and CT Findings of the 2019 Novel Coronavirus Disease (COVID-19): Analysis of Nine Patients Treated in Korea. Korean J Radiol. 2020;21(4):494-500. doi:10.3348/kjr.2020.0132

13. Xu X, Yu C, Qu J, et al. Imaging and clinical features of patients with 2019 novel coronavirus SARS-CoV-2. Eur J Nucl Med Mol Imaging. 2020;(613):1275-1280. doi:10.1007/s00259-020-04735-9

14. Albarello F, Pianura E, Di Stefano F, et al. 2019-novel Coronavirus severe adult respiratory distress syndrome in two cases in Italy: An uncommon radiological presentation. Int J Infect Dis. 2020;93:192-197. doi:10.1016/j.ijid.2020.02.043

15. Koo HJ, Lim S, Choe J, Choi SH, Sung H, Do KH. Radiographic and CT features of viral pneumonia. Radiographics. 2018;38(3):719-739. doi:10.1148/rg.2018170048

16. Simpson S, Kay FU, Abbara S, et al. Radiological Society of North America Expert Consensus Statement on Reporting Chest CT Findings Related to COVID-19. Endorsed by the Society of Thoracic Radiology, the American College of Radiology, and RSNA. Radiology: Cardiothoracic Imaging. Published Online:March 25 2020. doi: 10.1148/ryct.2020200152.

17. Jin Y, Cai L, Cheng Z, et al. A rapid advice guideline for the diagnosis and treatment of 2019 novel coronavirus (2019-nCoV) infected pneumonia (standard version). Mil Med Res. 2020;7(1):1-23. doi:10.1186/s40779-020-0233-6

18. Yuan M, Yin W, Tao Z, Tan W, Hu Y. Association of radiologic findings with mortality of patients infected with 2019 novel coronavirus in Wuhan, China. PLoS One. 2020;15(3):e0230548. doi:10.1371/journal.pone.0230548

19. Chaganti S, Balachandran A, Chabin G, et al. Quantification of Tomographic Patterns associated with COVID-19 from Chest CT. arXiv. 2020. Published Online April 2 2020. doi: eess.IV/2004.01279

20. Liu S, Georgescu B, Xu Z, et al. 3D Tomographic Pattern Synthesis for Enhancing the Quantification of COVID-19. arXiv. 2020. Published Online May 5 2020. doi: eess.IV/2005.01903

21. Shan F, Gao Y, Wang J, et al. Lung Infection Quantification of COVID-19 in CT Images with Deep Learning. arXiv e-prints. 2020. Published Online March 10 2020. doi: cs.CV/2003.04655

22. Narin A, Kaya C, Pamuk Z. Automatic Detection of Coronavirus Disease (COVID-19) Using X-ray Images and Deep Convolutional Neural Networks. arXiv. 2020. Published Online March 24 2020. doi: eess.IV/2003.10849

23. Ozturk T, Talo M, Yildrim E, et al. Automated detection of COVID-19 cases using deep neural networks with X-ray images. Computers in Biology and Medicine. 2020; 121: 103792.

24. Asnaoui K, Chawki Y. Using X-ray images and deep learning for automated detection of coronavirus disease. Journal of Biomolecular Structure and Dynamics. 2020; 0(0): 1-12. doi: 10.1080/07391102.2020.1767212

25. Yushkevich P, Gao Y, Gerig G. ITK-SNAP: an interactive tool for semi-automatic segmentation of multi-modality biomedical images. Conference proceedings: IEEE Engineering in Medicine and Biology Society. 2016: 3342-3345. doi: 10.1109/EMBC.2016.7591443.

26. Ronnenberger O, Fischer P, Brox, T. U-Net: Convolutional Networks for Biomedical Image Segmentation. arXiv. 2020. Published Online May 18 2015. doi: arXiv:1505.04597

27. Lakshminarayan B, Prizel A, Blundell C. Simple and Scalable Predictive Uncertainty Estimation using Deep Ensembles. arXiv. 2017. Published Online November 4 2017. doi: arXiv:1612.01474v3

28. ACR Recommendations for the use of Chest Radiography and Computed Tomography (CT) for Suspected COVID-19 Infection. https://www.acr.org/Advocacy-and-Economics/ACR-Position-Statements/Recommendations-for-Chest-Radiography-and-CT-for-Suspected-COVID19-Infection. Published Mar 11 2020.

29. Rubin GD, Ryerson CJ, Haramati LB, et al. The Role of Chest Imaging in Patient Management during the COVID-19 Pandemic: A Multinational Consensus Statement from the Fleischner Society. Radiology. 2020;296(1):172-180. doi:10.1148/radiol.2020201365

30. Chua F, Armstrong-James D, Desai S, et al. The role of CT in case ascertainment and management of COVID-19 pneumonia in the UK: insights from high-incidence regions. Respiratory Medicine. 2020; 8(5): 438-440. https://doi.org/10.1016/S2213-2600(20)30132-6

31. Li L, Qin L, Xu Z, et al. Artificial Intelligence Distinguishes COVID-19 from Community Acquired Pneumonia on Chest CT. Radiology. 2020. Published Online March 19 2020. doi: 10.1148/radiol.2020200905

32. Gozes O, Frid-Adar M, Greenspan H, et al. Rapid AI Development Cycle for the Coronavirus (COVID-19) Pandemic: Initial Results for Automated Detection & Patient Monitoring using Deep Learning CT Image Analysis. ArXiv. 2020. Published Online March 10 2020. doi: 2003.05037

33. Ying S, Shuangjia Z, Li L, et al. Deep learning Enables Accurate Diagnosis of Novel Coronavirus (COVID-19) with CT images. MedRxiv. 2020: 1-10. doi: 10.1101/2020.02.23.20026930

34. El Asnaoui K, Chawki Y. Using X-Ray images and deep learning for automated detection of coronavirus disease. Journal of Biomolecular Structure and Dynamics. 2020. Published Online May 22 2020. doi: 10.1080/07391102.2020.1767212

35. Kundu S, Elhalawani H, Gichoya J, et al. How Might AI and Chest Imaging Help Unravel COVID-19's Mysteries? Radiology: Artificial Intelligence. 2020; 2(3): e200053. doi: 10.1148/ryai.2020200053

**TABLES**

**Table 1: Demographics of Cohort (n=86)**

|  | Total (n=86) |
|---|---|
| **Age (mean, range)** | 59 (25-93) |
| **Gender** | |
| Male | 51% (44/86) |
| Female | 49% (42/86) |
| **Race/Ethnicity** | |
| White Non-Hispanic | 14% (12/86) |
| White Hispanic | 1% (1/86) |
| Black or African American | 75% (64/86) |
| Black Hispanic | 1% (1/86) |
| Asian | 2% (2/86) |
| East Indian | 1% (1/86) |
| Other/Unknown | 6% (5/86) |
| **Smoking Status** | |
| Never Smoked | 48% (41/86) |
| Former Smoker | 27% (23/86) |
| Current Smoker | 8% (7/86) |
| Unknown | 17% (15/86) |
| **Outcome** | |
| Discharged | 72% (62/86) |
| Deceased | 23% (20/86) |
| Still admitted (as of 7/13/2020) | 5% (4/86) |

**Table 2: CT and CXR characteristics of the cohort regarding distribution of POv and POa.**

| | Reader POv / POa stats. | | | | |
|---|---|---|---|---|---|
| | Modality | min | max | mean | standard deviation |
| Ground-Truth POv | CT | 0 | 0.97 | 0.23 | 0.21 |
| DRR POa | DRR | 0 | 0.98 | 0.19 | 0.24 |

**Table 3: 95% bootstrapped mean absolute errors (MAE) and Pearson coefficients of the CNN systems and the expert readers on CXR against the ground truth of POv measured on CT.**

| Description | Modality | 95% Bootstrap MAE | 95% Bootstrap Pearson |
|---|---|---|---|
| Reader Avg. CXR | CXR | 11.98% (11.05% to 12.47%) | 0.77 (0.70 to 0.82) |
| Reader Inter. CXR | CXR | 15.91% (14.68% to 16.63%) | 0.73 (0.66 to 0.78) |
| Reader Union CXR | CXR | 11.04% (10.22% to 11.50%) | 0.77 (0.69 to 0.82) |
| Single CNN DRR | DRR | 9.27% (8.56% to 9.67%) | 0.86 (0.84 to 0.90) |
| Ensemble CNN DRR | DRR | 7.73% (6.74% to 8.07%) | 0.87 (0.84 to 0.92) |
| Single CNN CXR | CXR | 9.78% (8.94% to 10.22%) | 0.78 (0.73 to 0.82) |
| Ensemble CNN CXR | CXR | 9.56% (8.83% to 10.00%) | 0.81 (0.76 to 0.85) |

**Table 4: Statistical comparison of readers (average, intersection, union) versus CNNs (single, ensemble) for quantification of POa on CXR. t-scores and the one-tail p-values from the related sample t-test.**

| | t-score | p-value |
|---|---|---|
| **Reader Avg vs. Single CNN** | 1.68 | 0.05 |
| **Reader Avg vs. Ensemble CNN** | 2.46 | 0.01* |
| **Reader Intersect vs. Single CNN** | 3.71 | <0.01* |
| **Reader Intersect vs. Ensemble CNN** | 4.68 | <0.01* |
| **Reader Union vs. Single CNN** | 1.31 | 0.10 |
| **Reader Union vs. Ensemble CNN** | 1.90 | 0.03* |
| **Single CNN vs. Ensemble CNN** | 0.38 | 0.35 |

**FIGURES**

**Figure 1: Flow diagram with inclusion and exclusion criteria in our cohort (n = 86)**

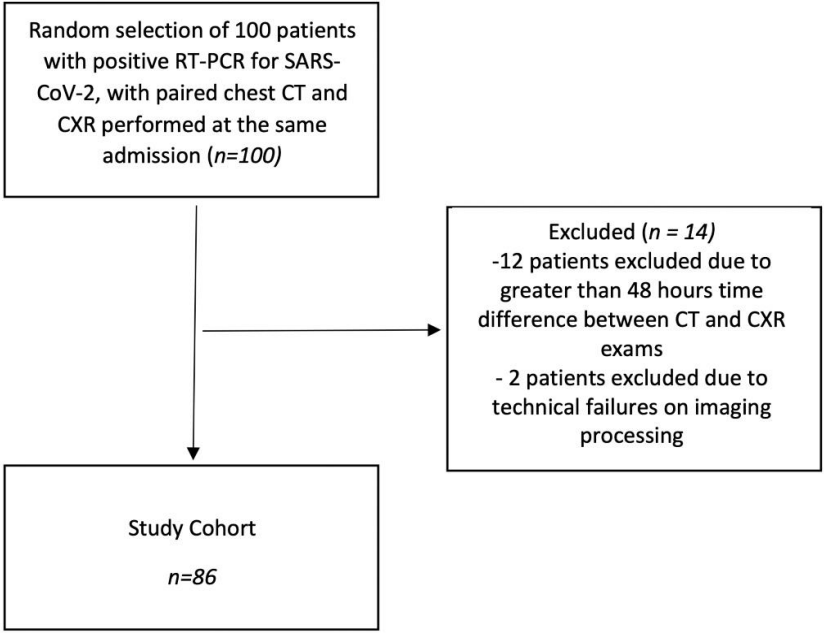

**Figure 2: Schematic illustration of the data-flow for 3D AD POv quantification and projection on super resolution up-sampled DRR for POa quantification based on CT derived 3D AD POv mask.**

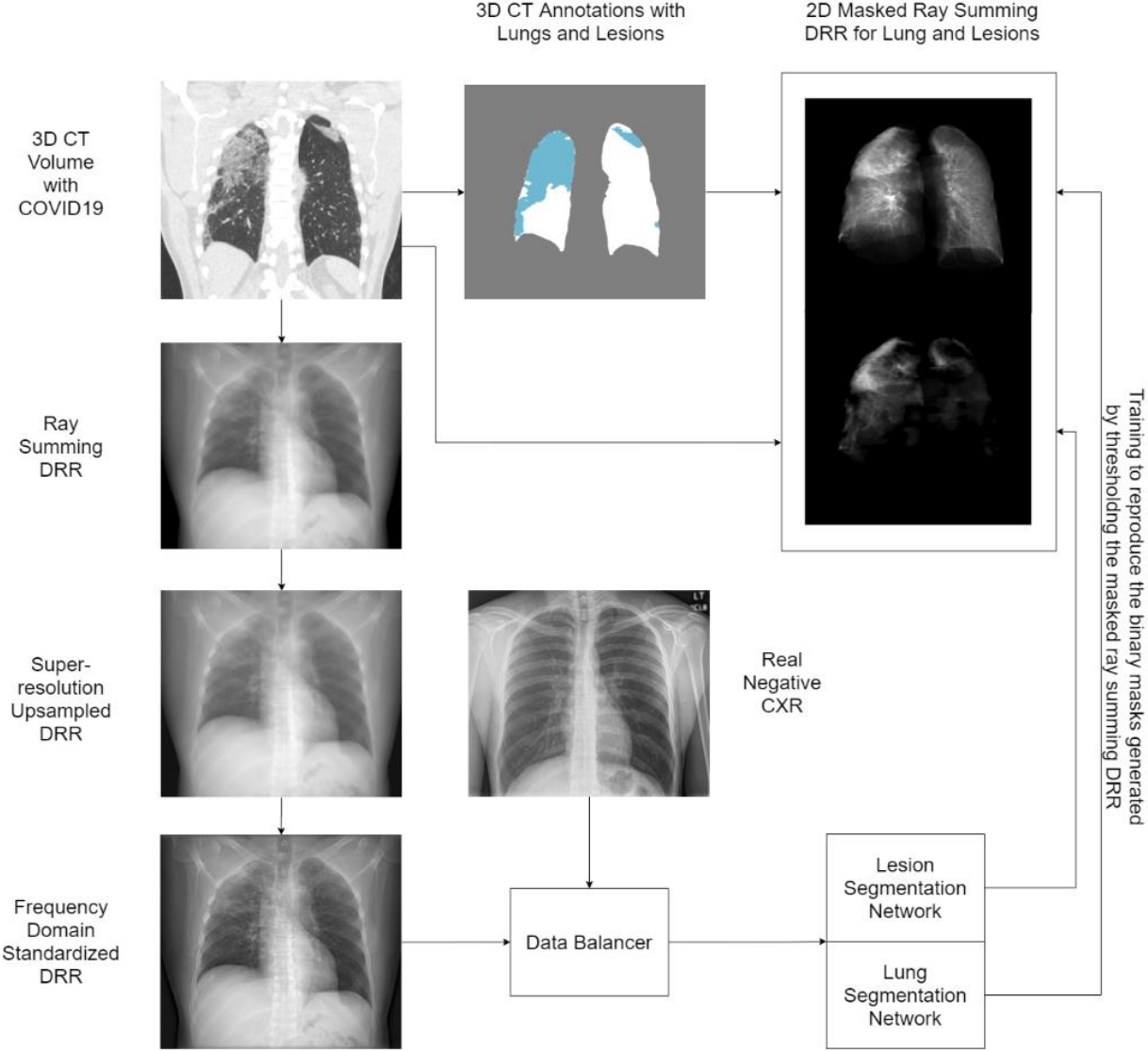

**Figure 3:** Schematic illustration of the study design with comparison of expert reader annotation on CXR, CNN prediction on CXR, CNN prediction on DRR, DRR POa derived from CT projection, and CT derived POv (ground-truth).

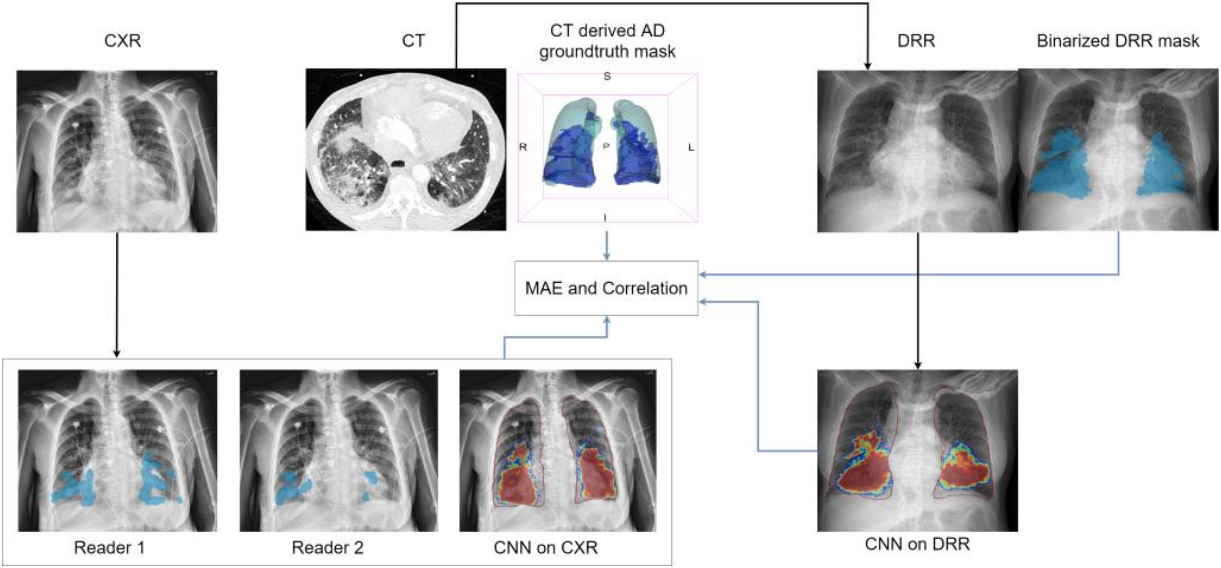

**Figure 4:** CT derived 3D volume (POv) airspace quantification (ground truth), on (A, E) axial, (B, F) sagittal, (C, G) coronal MPRs and on (D, H) VR mask, for 2 patients in our cohort

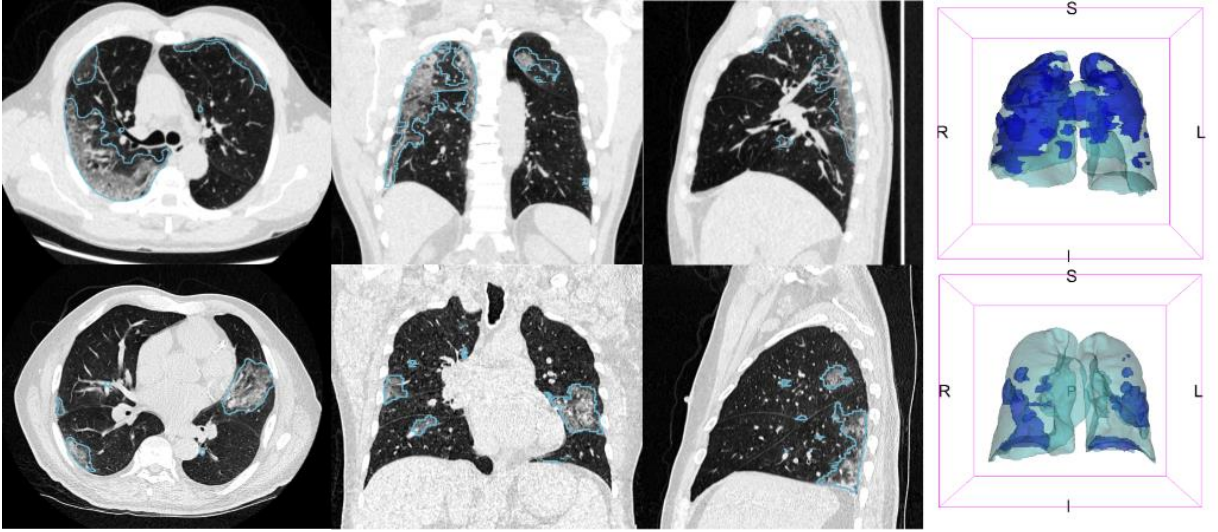

**Figure 5: DRR (A), intensity/AP thickness AD mask from CT (B), DRR + AD mask (C), DRR CNN prediction (D), CXR (E), CXR reader 1 (F), CXR reader 2 (G), CXR CNN prediction (H), for a patient example in our cohort**

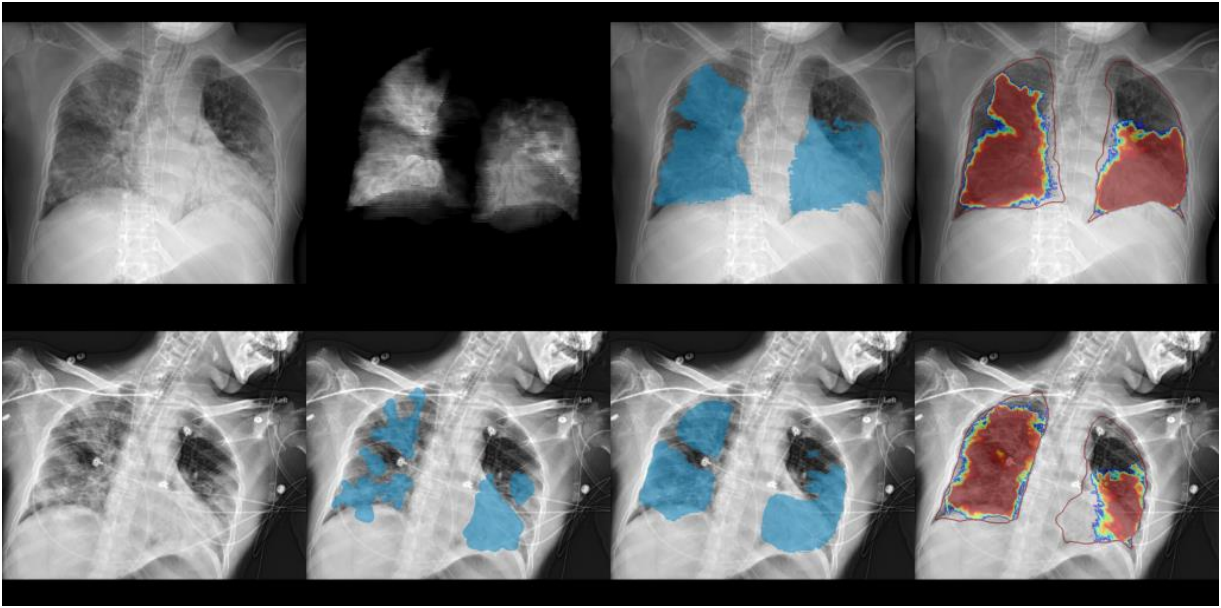

**Figure 6: DRR (A), intensity/AP thickness AD mask from CT (B), DRR + AD mask (C), DRR CNN prediction (D), CXR (E), CXR reader 1 (F), CXR reader 2 (G), CXR CNN prediction (H), for another patient example in our cohort**

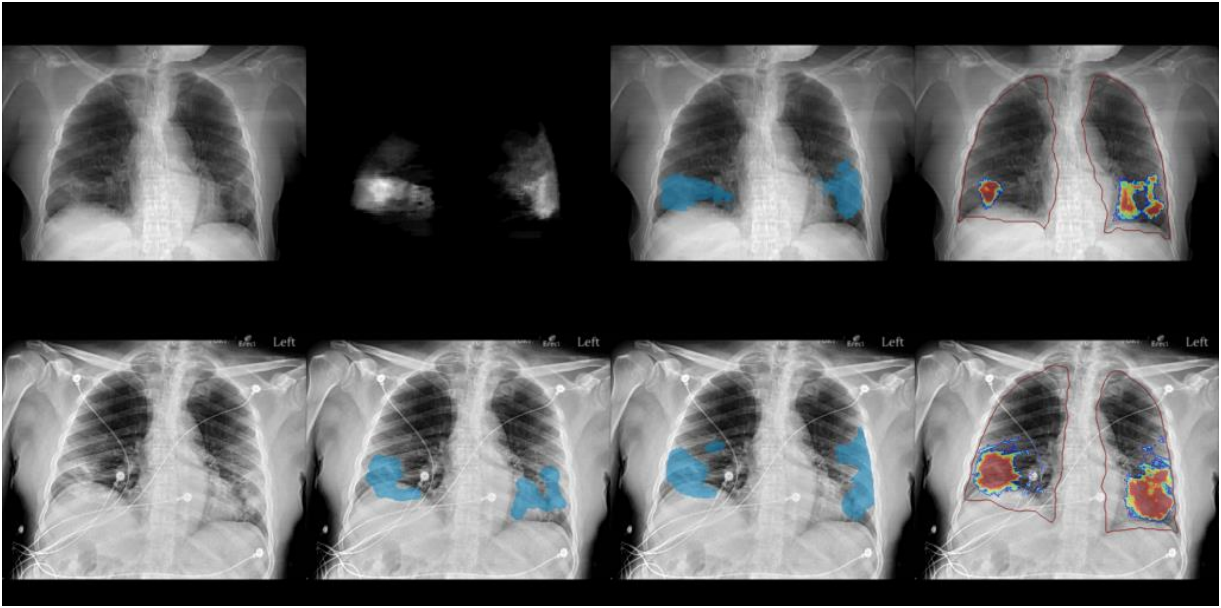

Supplemental Table: **Properties of the Training and Validation Data used for Development of the AI Lung and Airspace Disease Segmentation on CT and CXR, compared to our study cohort.**

|  | AI System Training CT | AI system Training CXR | AI System Validation CT | Study Cohort CT |
|---|---|---|---|---|
| Datasets | Total: 1929, COVID-19: 1005, ILD: 267, Pneumonia: 147, Normal: 510, | Total: 727 COVID-19: 0 Control: 727 | Total: 182, COVID-19: 86, Control: 96 | Total: 86, COVID-19: 86 |
| Data Origin | Multiple sites including USA, Spain, Switzerland, Germany, France, Denmark, Canada and Belarus, excluding Upenn | Multiple sites including USA and Germany, excluding Upenn | Multiple sites including USA, Spain and Czech Republic, excluding Upenn | Upenn |
| Sex | Female: 628, Male: 827, Unknown: 474 | Female: 142 Male: 123 Unknown: 461 | Female: 66, Male: 101, Unknown: 15 | Female: 42, Male: 44, Unknown: 0 |
| Age (years) | Median: 61 IQR: 56-66 Unknown: 938 | Median: 58 IQR: 31 Unknown: 461 | Median: 62 IQR: 53-72 Unknown: 54 | Median: 59 IQR : 25-93 Unknown: 0 |
| Scanner Manufacturer | GE: 450, Siemens: 1258, Philips: 41, Toshiba: 23, Other/Unknown: 156 | 'FUJIFILM': 111, 'Carestream': 130, 'Agfa': 25, Other/Unknown: 461 | GE: 60, Siemens: 58, Philips: 24, Toshiba: 27, Other/Unknown: 13 | GE: 29, Siemens: 57, Philips: 0, Toshiba: 0 Other/Unknown: 0 |
| Slice Thickness [mm] | ≤ 1.5: 1632 (1.5, 3.0]: 282, >3.0: 12 | N/A | ≤ 1.5: 51, (1.5, 3.0]: 116, >3.0: 15 | ≤ 1.5: 82, (1.5, 3.0]: 4, >3.0: 0 Unknown: 0 |
| Reconstruction Kernel | Soft: 691 Hard: 1035 Unknown 203 | N/A | Soft: 86 Hard: 71 Unknown: 25 | Soft: 32 Hard: 54 Unknown: 0 |